\documentclass[cup6b]{cupbook}
\usepackage{graphicx}
\usepackage{natbib}
\title[Rome, Italy, 27--30 April 2009]
      {The coming of age of X-ray polarimetry}
\author{}
\date{}
\begin{document}
\pagenumbering{arabic}

\def\arcmin{\hbox{$^\prime$}}
\def\arcsec{\hbox{$^{\prime\prime}$}}
\def\deg{\hbox{$^\circ$}}
\def\sun{\hbox{$\odot$}}
\def\earth{\hbox{$\oplus$}}
\def\lae{\mathrel{\raise .4ex\hbox{\rlap{$<$}\lower 1.2ex\hbox{$\sim$}}}}
\def\gae{\mathrel{\raise .4ex\hbox{\rlap{$>$}\lower 1.2ex\hbox{$\sim$}}}}
\let\lsim=\lae
\let\gsim=\gae

\def\aap{{A\&A}}
\def\aaps{{A\&AS}}
\def\aj{{AJ}}
\def\apj{{ApJ}}
\def\apjl{{ApJ}}
\def\apjs{{ApJS}}
\def\ao{{Appl.Optics}}
\def\mnras{{MNRAS}}
\def\nat{{Nature}}
\def\pasj{{PASJ}}
\def\pasp{{PASP}}

\author[Marshall et al.]{Herman L. Marshall, Ralf Heilmann,  Norbert Schulz \and Kendrah Murphy (MIT Kavli Institute)}
\chapter{Broad-Band Soft X-ray Polarimetry}

\abstract{We developed an instrument design capable of measuring linear X-ray polarization
over a broad-band using conventional spectroscopic optics.
A set of multilayer-coated ßats reßects the dispersed X-rays to the instrument detectors.
The intensity variation with position angle is measured to determine
three Stokes parameters: I, Q, and U -- all as a function of energy.
By laterally grading the multilayer optics and matching the dispersion of the gratings,
one may take advantage of high multilayer reßectivities and achieve
modulation factors $>$50\% over the entire 0.2 to 0.8 keV band.
This instrument could be used in a small orbiting mission or scaled up for the International X-ray Observatory.
Laboratory work has begun that would demonstrate the capabilities of key components.}

\section{Introduction}

The soft X-ray band (0.1-1.0 keV) should prove to be a fruitful region to
explore for polarized emission.
One concept, the Polarimeter for Low Energy X-ray Astrophysical
Sources (PLEXAS), was proposed to use
multilayer-coated mirrors tuned to 0.25 keV as Bragg reflectors\citep{plexas}.
As in similar Bragg reflection systems, the PLEXAS
design had a narrow bandpass, reducing its attractiveness
for astrophysical observations because one expects polarization to be energy
dependent, so a wide bandpass is desired.

Marshall (2007\citep{2007SPIE.6688E..31M})
described a method to overcome this limitation by using transmission gratings
to disperse in the incoming X-rays.
Following up on this approach,
Marshall (2008\citep{2008SPIE.7011E..63M}) suggested
an arrangement that can be used in missions ranging from
a small explorer to the International X-ray Observatory (IXO).

\section{Polarimetry with the Extreme Ultraviolet Explorer}

For the record, it is worth noting that the first attempt to measure
the polarization above 0.1 keV of an extragalactic source used
the Extreme Ultraviolet Explorer (EUVE).
However, the telescope was not designed for polarimetry.
The expected modulation factor was small, 2.7\%, so
experiment failed due to systematic errors that could not
be eliminated with existing calibration data.\citep{pkspolar}
Even using an in-flight null calibrator (a white dwarf) was
insufficient.  The program described here grew out of this
attempt, so one might say that it was not a complete loss.

\section{Science Goals}

Here we describe two potential scientific studies to be
performed with an X-ray polarimetry mission with sensitivity
in the 0.1-1.0 keV band.

{\bf Probing the Relativistic Jets in BL Lac Objects --}
\label{bllscience}
Blazars are all believed to contain parsec-scale jets with
$\beta \equiv v/c$ approaching 0.995.
The jet and shock
models make different predictions regarding the directionality of the
magnetic field at X-ray energies: for knots in a laminar jet flow it should lie nearly
parallel to the jet axis \citep{1980ApJ...235..386M}, while for shocks it should lie
perpendicular \citep{1985ApJ...298..114M}.
McNamara et al. (these proceedings) suggest that
X-ray polarization data could be used to deduce the primary emission mechanism
at the base, discriminating between synchrotron, self-Compton (SSC), and
external Compton models.  Their SSC models predict polarizations between
20\% and 80\%, depending on the uniformity of seed photons and the inclination angle\citep{2009MNRAS.395.1507M}.
The X-ray spectra are usually very steep so that a small instrument operating
below 1 keV can be quite effective.

{\bf Polarization in Disks of Active Galactic Nuclei and X-ray Binaries --}
Schnittmann \& Krolik (these proceedings)
show that the variation of polarization with energy could be used
as a probe of the black hole spin and that the polarization position
angle would rotate through $90\deg$ between 1 and 2 keV in some
cases, arguing that X-ray polarization measurements are
needed both below and above 2 keV\cite{2009arXiv0902.3982S}.
As Blandford et al. (2002) noted ``to
understand the inner disk we need ultraviolet and X-ray polarimetry''
\citep{2002apsp.conf..177B}.

\section{Basis of a Soft X-ray Polarimeter}

The approach to this polarimeter design was inspired by a new blazed transmission grating
design, called the Critical Angle Transmission (CAT) grating \citep{Heilmann08}, and the
corresponding application to the Con-X mission in the design of
a transmission grating spectrometer\citep{flanagan07}.
This type
of grating can provide very high efficiency in first order in the soft
X-ray band.  For a spectrometer, one places detectors on the Rowland
torus, which is slightly ahead of the telescope's imaging surface.

\begin{figure}
   \centering
   \includegraphics*[height=9cm,angle=0]{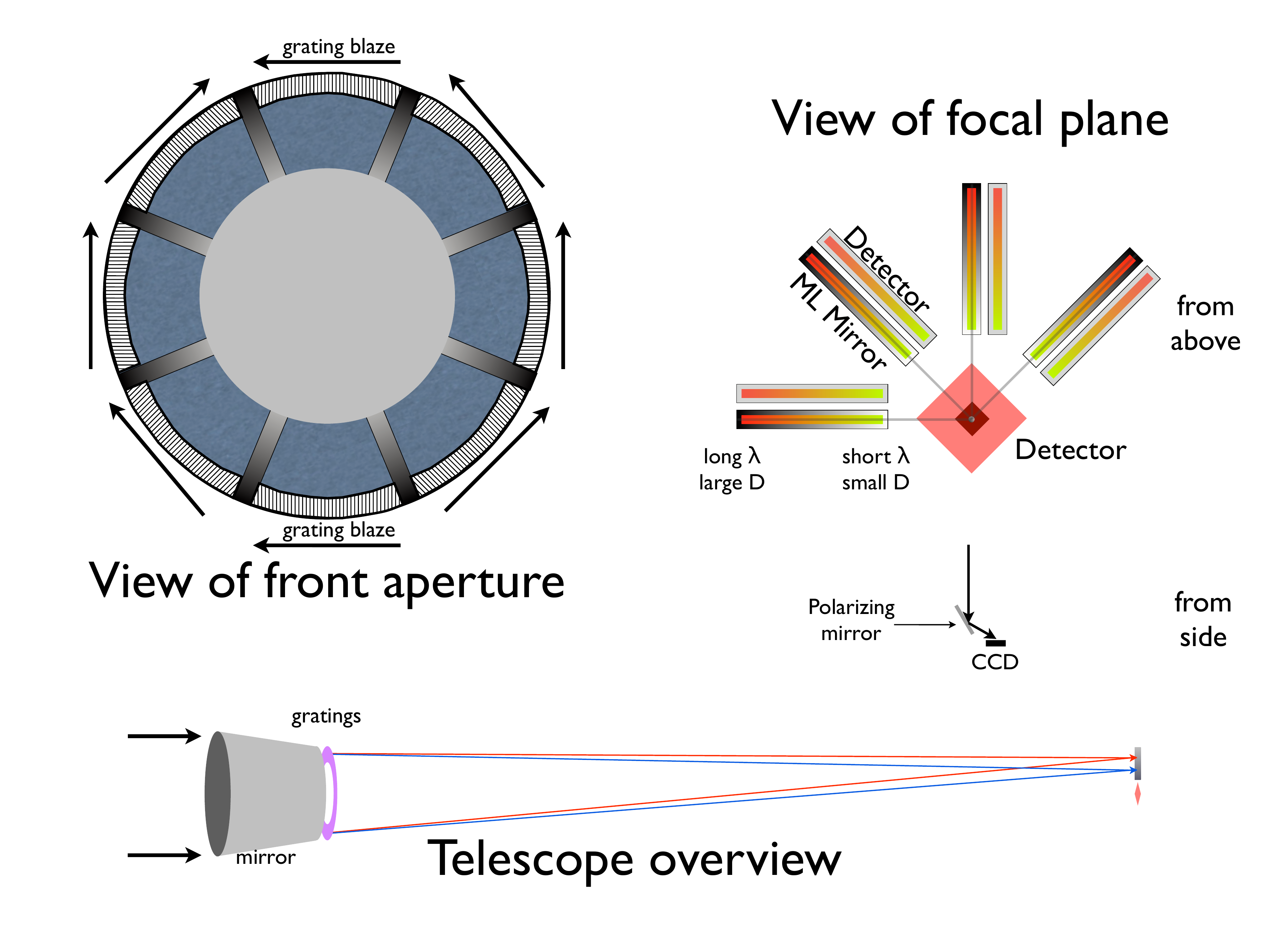}
 \caption{
{\it Bottom:} Overview of a small telescope designed for soft X-ray spectropolarimetry,
based on a design suggested by Marshall (2008, \cite{2008SPIE.7011E..63M}).
A small set of nested mirror shells focusses X-rays through gratings that disperse to
an array of detectors about 2 m from the entrance aperture.
{\it Top left:} View from the front aperture.
Gratings are placed
behind the mirror with the grating bars oriented along the average radius to the mirror axis.
The gratings are blazed in the directions shown.
 {\it Top right:} The view from above the detectors and polarizers.  Spectra from the gratings are
 incident on multilayer-coated flats that are tilted along the dispersion axis which contains
 the zeroth order (dot in center).  The angle of the tilt is the same for all mirrors and always
 redirects the X-rays to the adjacent detector in a clockwise direction.  The multilayer
 coating spacing, $D$, increases linearly outward from zeroth order, just as the wavelength
 increases, in order to match the first order wavelength to the peak of the multilayer reflectivity.
The detectors could be aligned to the appropriate Rowland torus (as shown
in the side view) or be tilted to face
the corresponding mirror.  Measuring the intensity at a given wavelength as a function
of clocking angle then provides $I(\lambda)$, $Q(\lambda)$, and $U(\lambda)$.
 }
\label{fig:layoutsmex}
\end{figure}

A dispersive spectrometer becomes a polarimeter
by placing multilayer-coated flats on the Rowland circle that
redirect and polarize the spectra.  The flats are tilted about
the spectral dispersion axis by an angle $\theta$.
For this design\cite{2008SPIE.7011E..63M}, the graze angle $\theta$ was
chosen to be $40\deg$.
The detectors are then
oriented toward the mirrors at an angle $90 - 2 \theta = 10\deg$ to the
plane perpendicular to the optical axis of the telescope.
Fig.~\ref{fig:layoutsmex} shows how the optics might look.
In this case, the entrance aperture is divided into eight sectors,
with gratings aligned to each other in each sector and along the
average radial direction to the optical axis.  This approach
generates seven spectra that are reflected by four polarizing
flats to four detectors.  See Marshall (2008) for details\cite{2008SPIE.7011E..63M}.

The fundamental requirement of this approach
is to vary the multilayer spacing, $d$, on the polarizing flats
laterally, along the dispersion, in order to provide optimal reflection
at graze angle $\theta$.
The multilayer period varies linearly with $x$, providing high reflectivity
in a narrow bandpass at large graze angles\cite{2008SPIE.7011E..63M}.
At Brewster's angle, $\theta = 45\deg$, reflectivity is minimized when the $E$-vector
is in the plane containing the incident ray and the surface normal ($p$-polarized)
and maximized when the $E$-vector is in the surface plane ($s$-polarized).
The polarization position angle (PA) is the average orientation in sky coordinates
of the $E$-vector for the incoming X-rays.
Sampling at least 3 PAs is required in order to measure
three Stokes parameters (I, Q, U) uniquely, so one would require at
least three separate detector
systems with accompanying multilayer-coated flats.
For this study, four detector-flat combinations
are assumed, as shown in figure~\ref{fig:layoutsmex}.

Critical angle transmission (CAT) gratings
have excellent prospects for high efficiency in first order,
up to 50\% over a wide band.\citep{Heilmann08}
The CAT gratings are free-standing (so no support membrane is needed)
and have higher intrinsic efficiency than the gratings used on the {\em Chandra} program,
making them excellent candidates for use in X-ray spectrometers as well as a possible X-ray
polarimeter.

 \begin{figure}
   \centering
   \includegraphics[height=8cm]{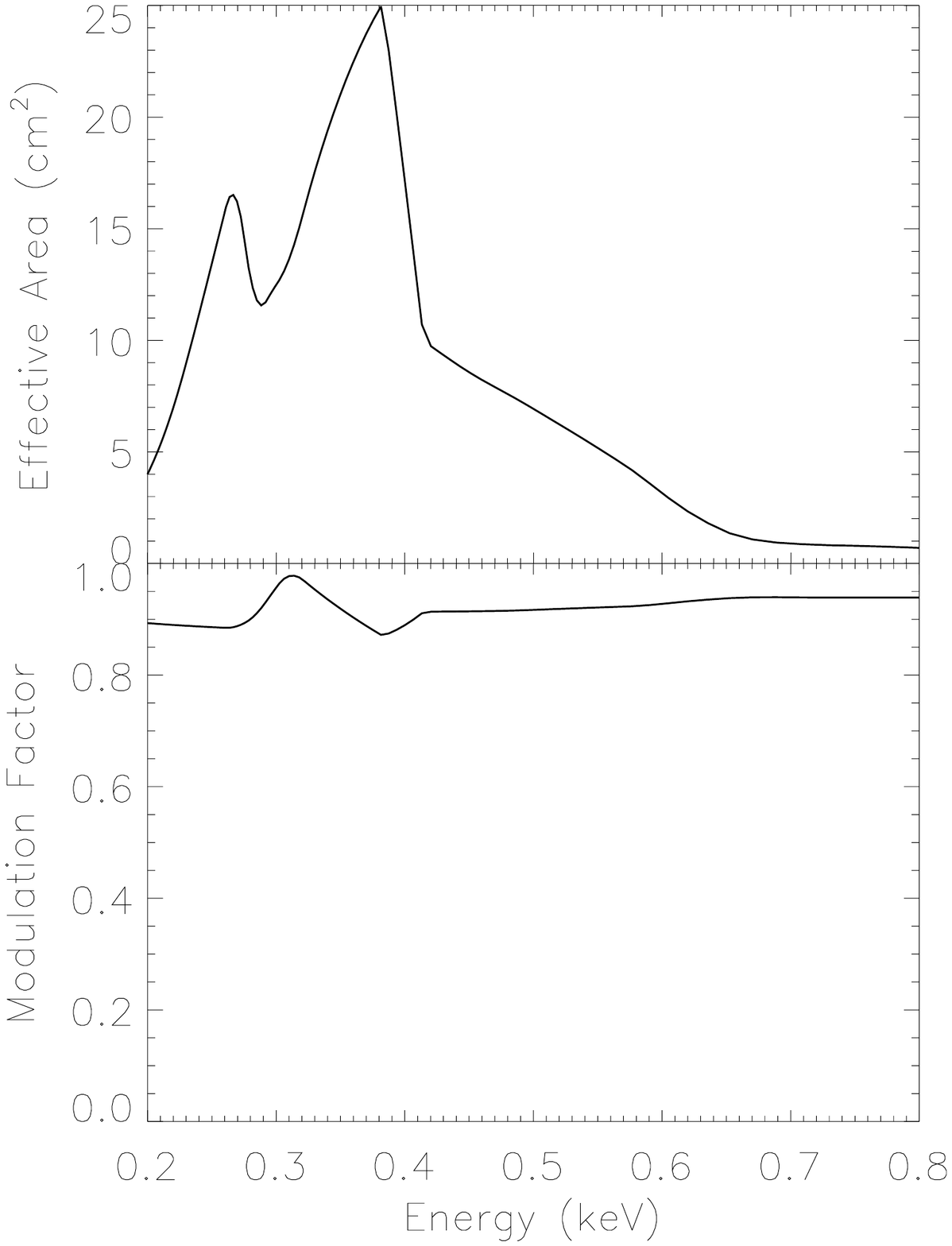}
   \includegraphics[height=6cm,angle=90]{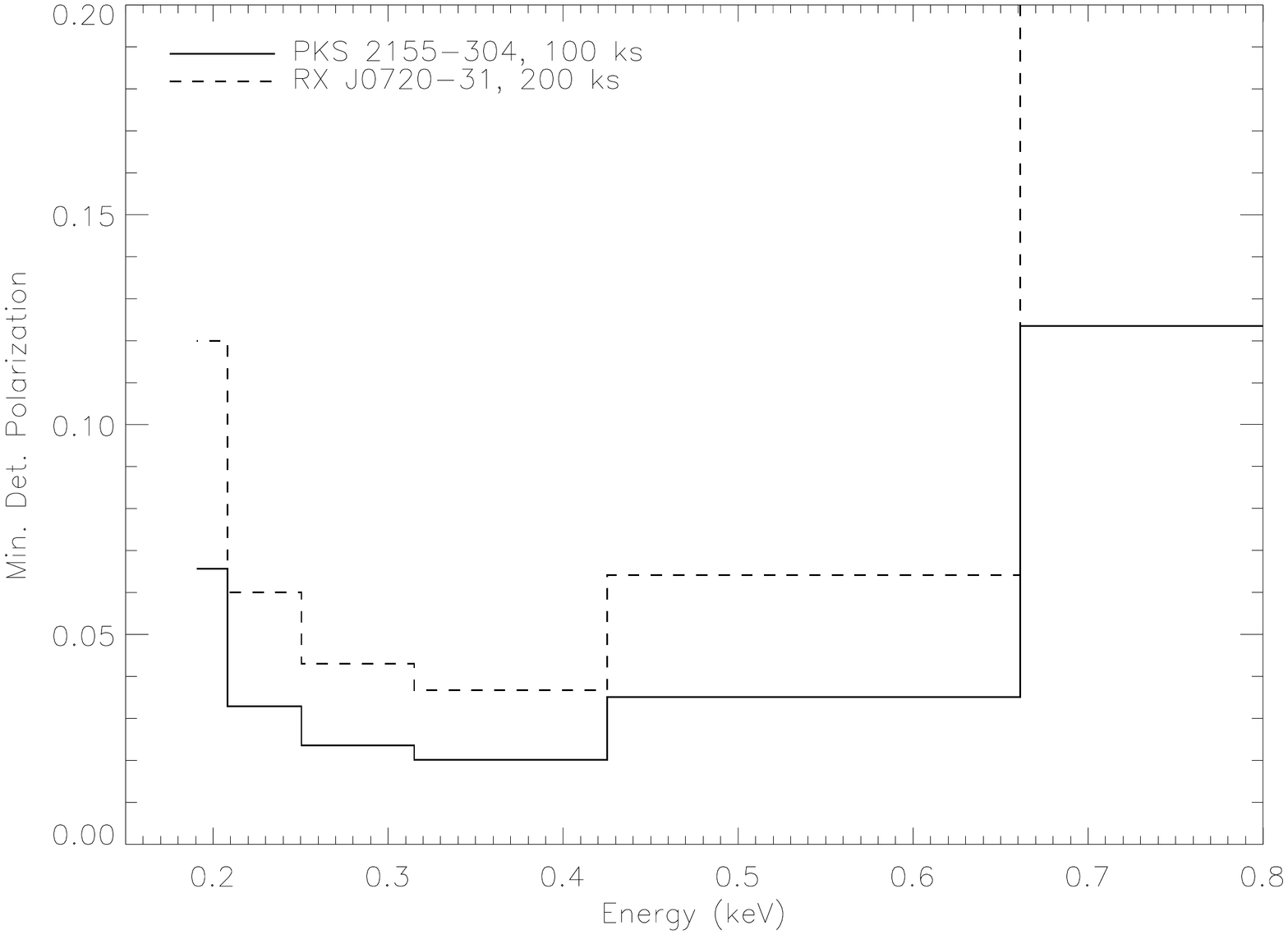}
 \caption{
{\it Left:} The effective area of a small mission, to unpolarized light.
	The geometric area of the broad-band mirror is assumed to be
	200 cm$^2$.  For details, see Marshall (2008, \cite{2008SPIE.7011E..63M}).
{\it Right:} Minimum detectable polarization as a function of energy across the
bandpass of the instrument for two different possible observations.
The solid line shows how we could detect linear polarization at a level
of 15-20\% across the
entire energy band from 0.2 to 0.8 keV for PKS 2155-304 in 100 ks.
For RX J0720-31, spectroscopy allows one to obtain the polarization
below, in, and above absorption features.
}
\label{fig:smex}
\end{figure}

\begin{figure}
    \centering
   \includegraphics[height=9cm]{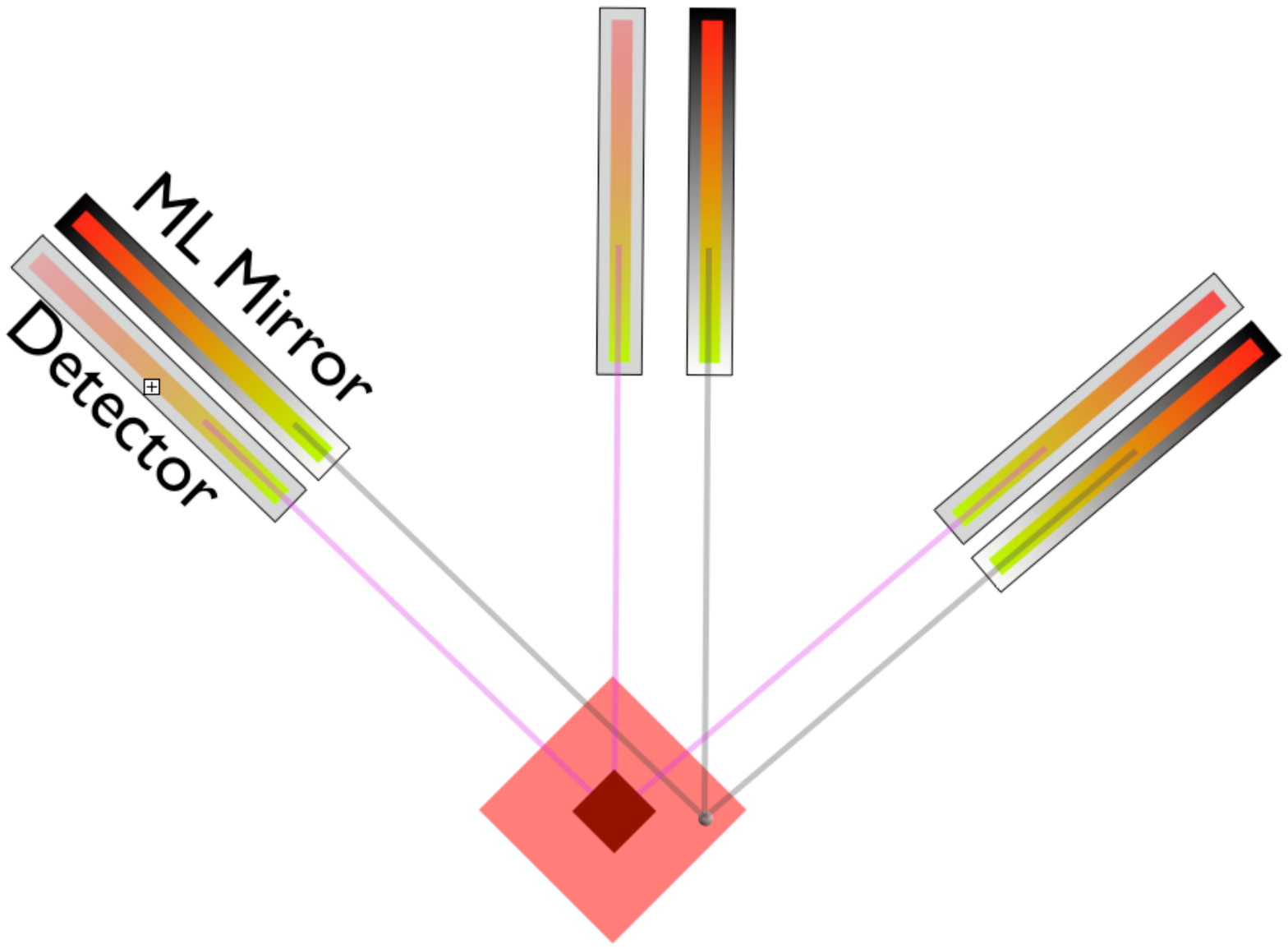}
   \includegraphics[height=6cm]{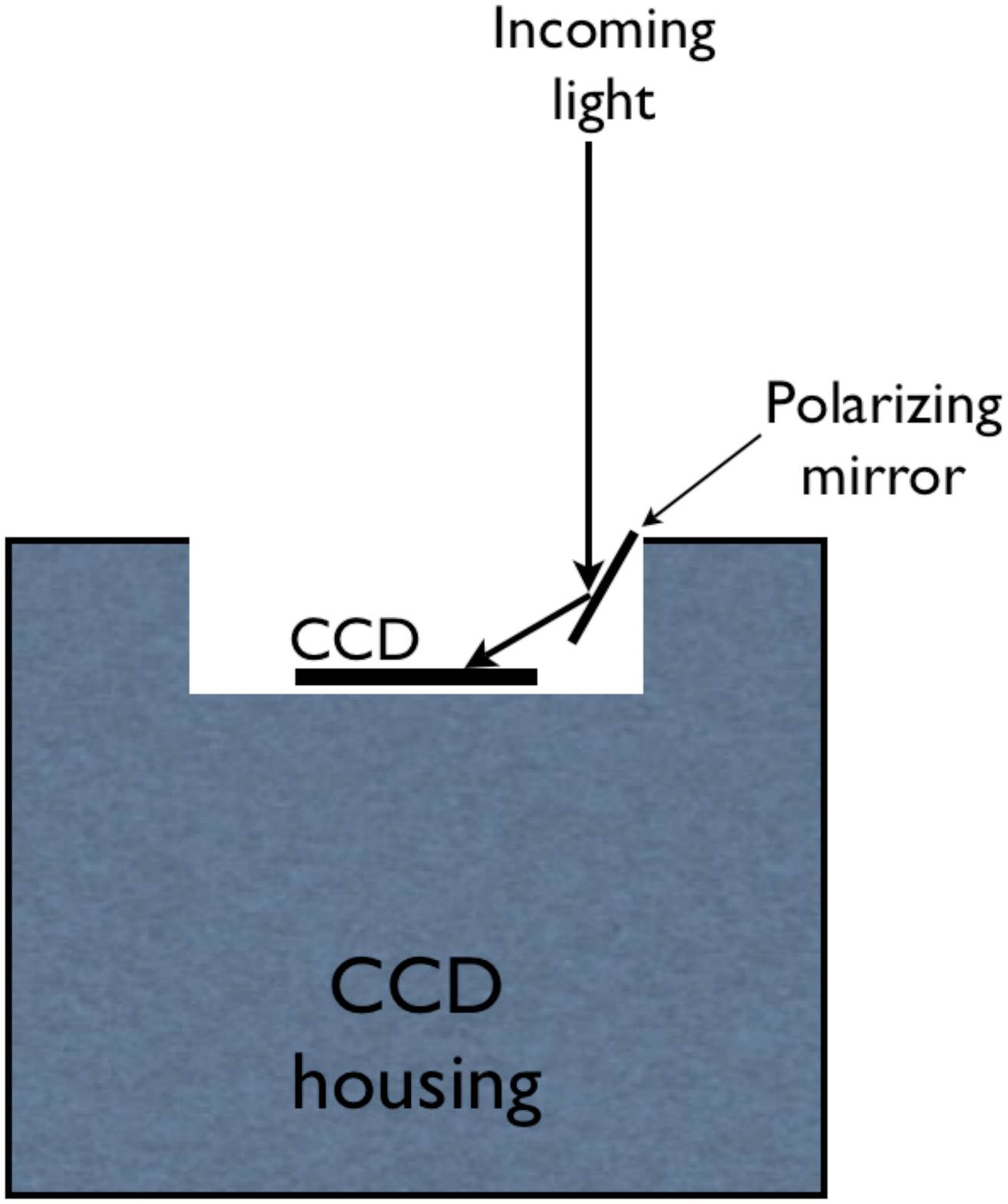}
 \caption{
{\it Left:} Top view of a focal plane layout that could be used for IXO, in the manner suggested by
	Marshall (2008, \cite{2008SPIE.7011E..63M}).  When used as a spectrometer (pink lines), the zeroth order
	of the spectrometer is centered in the dark red square, representing an IXO wide-field imager.
	When used as a polarimeter (gray lines), the zeroth order is placed at the location of the gray dot so that
	the dispersed spectrum first intercepts the laterally graded multilayer mirror that is angled at
	$30\deg$ to the incoming X-rays.  The lines leading to zeroth order are all the same length,
	showing that specific wavelengths appear at different distances from the near end of the
	mirror but are at the same distances when used as a spectrometer.
	{\it Right:} Side view of the CCD housing where the dispersion is perpendicular to the plane
	of the drawing and the multilayer mirror is oriented $30\deg$ to the incoming, dispersed X-rays.
}
\label{fig:ixo}
\end{figure}

A CAT grating spectrometer has been proposed as one of two approaches to use
on the IXO, so there is a development path for this type of soft X-ray polarimeter.
Sampling at least 3 PAs is required in order to measure
three Stokes parameters (I, Q, U) uniquely, so one would require at
least three separate detector
systems with accompanying multilayer-coated flats, as shown in fig.~\ref{fig:ixo}.

\section{A Soft X-ray Polarimeter Prototyping Facility}

We have recently recommissioned the
X-ray grating evaluation facility (X-GEF), a 17 m beamline that was developed
for testing transmission gratings fabricated at MIT for
the {\it Chandra} project\citep{1994SPIE.2280..257D}.
With MKI technology development funding, we will soon
adapt the existing source to produce polarized X-rays at the O-K$\alpha$
line (0.525 keV).
A new chamber will house a Polarized Source MultiLayer mirror,
provided by Reflective X-ray Optics (RXO).  The source will be mounted in the chamber
at $90\deg$ to the existing beamline.  A rotatable flange will connect the
PSML chamber output to the vacuum 17m facility so that the polarization
vector may be rotated with respect to the grating and its dispersion direction.
This prototyping effort will help us demonstrate the viability
of soft X-ray polarimetry and show how a broad-band version can
be developed.  A proposal to expand the facility to 4 or 5 more energies
between 0.25 and 0.8 keV is under consideration by NASA.


\begin{thereferences}{99}

\bibitem{2002apsp.conf..177B} Blandford, R., et al.\ (2002). 
	In \textit{Astrophysical Spectropolarimetry}, 
	ed. J.~{Trujillo-Bueno}, F.~{Moreno-Insertis}, and F.~{S{\'a}nchez}.

\bibitem{1994SPIE.2280..257D} Dewey, D., et al.\ (1994).
	\textit{SPIE} \textbf{2280}, 257--271.\  

\bibitem{flanagan07} Flanagan, K., et al.\ (2007).
	\textit{SPIE} \textbf{6688}, 66880Y.

\bibitem{Heilmann08} Heilmann, R.~K., et al.\ (2008). 
	\textit{Opt. Express} \textbf{16}, 8658--8669.

\bibitem{1980ApJ...235..386M} Marscher, A.~P.\ (1980).
	\textit{Ap. J.} \textbf{235}, 386--391.\  

\bibitem{1985ApJ...298..114M} Marscher, A.~P., \& Gear, W.~K.\ (1985).
	\textit{ApJ} \textbf{298}, 114--127.\  

\bibitem{2008SPIE.7011E..63M} Marshall, H.~L.\ (2008).
	\textit{SPIE} \textbf{7011}, 701129.

\bibitem{2007SPIE.6688E..31M} Marshall, H.~L.\ (2007).
	\textit{SPIE} \textbf{6688}, 66880Z.

\bibitem{plexas} Marshall, H.~L., et al.\ (2003). 
	\textit{SPIE} \textbf{4843}, 360--371.\  

\bibitem{pkspolar} Marshall, H.~L., et al.\ (2001).
	\textit{Ap. J.} \textbf{549}, 938--947.\  

\bibitem{2009MNRAS.395.1507M} McNamara, A.~L., et al.\ (2009). 
	\textit{MNRAS} \textbf{395}, 1507--1514.\  

\bibitem{2009arXiv0902.3982S} Schnittman, J.~D., 	\& Krolik, J.~H.\ (2009).
	\textit{arXiv:0902.3982}.

\end{thereferences}
\end{document}